\documentclass[twocolumn,prl,aps,superscriptaddress,showpacs,floatfix,lengthcheck]{revtex4}
\usepackage{graphicx}
\usepackage{color}
\begin{document}
\title{The proton elastic form factor ratio $\mu_{p} G_E^p/G_M^p$ at low momentum transfer}
\author{G.~Ron}
\affiliation{Tel Aviv University, Tel Aviv 69978, Israel} 
\author{J.~Glister}
\affiliation{Saint Mary's University, Halifax, Nova Scotia B3H 3C3, Canada}
\affiliation{Dalhousie University, Halifax, Nova Scotia B3H 3J5, Canada}
\author{B.~Lee}
\affiliation{Seoul National University, Seoul 151-747, Korea}
% Collaboration
\author{K.~Allada}
\affiliation{University of Kentucky, Lexington, Kentucky 40506, USA}
\author{W.~Armstrong}
\affiliation{Temple University, Philadelphia, Pennsylvania 19122, USA}
\author{J.~Arrington}
\affiliation{Argonne National Laboratory, Argonne, Illinois 60439, USA}
\author{A.~Beck}
\affiliation{NRCN, P.O.Box 9001, Beer-Sheva 84190, Israel}
\author{F.~Benmokhtar}
\affiliation{University of Maryland, Baltimore, Maryland, USA}
\author{B.L.~Berman}
\affiliation{George Washington University, Washington D.C. 20052, USA}
\author{W.~Boeglin}
\affiliation{Florida International University, Miami, Florida 33199, USA}
\author{E.~Brash}
\affiliation{Christopher Newport University, Newport News, Virginia, 2360X, USA}
\author{A.~Camsonne}
\affiliation{Thomas Jefferson National Accelerator Facility, Newport News, Virginia 23606, USA}
\author{J.~Calarco}
\affiliation{University of New Hampshire, Durham, New Hampshire 03824, USA}
\author{J.~P.~Chen}
\affiliation{Thomas Jefferson National Accelerator Facility, Newport News, Virginia 23606, USA}
\author{Seonho~Choi}
\affiliation{Seoul National University, Seoul 151-747, Korea}
\author{E.~Chudakov}
\affiliation{Thomas Jefferson National Accelerator Facility, Newport News, Virginia 23606, USA}
\author{L.~Coman}
\affiliation{}
\author{B.~Craver}
\affiliation{University of Virginia, Charlottesville, Virginia 22904, USA}
\author{F.~Cusanno}
\affiliation{INFN, Sezione Sanit\'{a} and Istituto Superiore di Sanit\'{a}, 
Laboratorio di Fisica, I-00161 Rome, Italy}
\author{J.~Dumas}
\affiliation{Rutgers, The State University of New Jersey, Piscataway, New Jersey 08855, USA}
\author{C.~Dutta}
\affiliation{University of Kentucky, Lexington, Kentucky 40506, USA}
\author{R.~Feuerbach}
\affiliation{Thomas Jefferson National Accelerator Facility, Newport News, Virginia 23606, USA}
\author{A.~Freyberger}
\affiliation{Thomas Jefferson National Accelerator Facility, Newport News, Virginia 23606, USA}
\author{S.~Frullani}
\affiliation{INFN, Sezione Sanit\'{a} and Istituto Superiore di Sanit\'{a}, 
Laboratorio di Fisica, I-00161 Rome, Italy}
\author{F.~Garibaldi}
\affiliation{INFN, Sezione Sanit\'{a} and Istituto Superiore di Sanit\'{a}, 
Laboratorio di Fisica, I-00161 Rome, Italy}
\author{R.~Gilman}
\affiliation{Thomas Jefferson National Accelerator Facility, Newport News, Virginia 23606, USA}
\affiliation{Rutgers, The State University of New Jersey, Piscataway, New Jersey 08855, USA}
\author{O.~Hansen}
\affiliation{Thomas Jefferson National Accelerator Facility, Newport News, Virginia 23606, USA}
\author{D.~W.~Higinbotham}
\affiliation{Thomas Jefferson National Accelerator Facility, Newport News, Virginia 23606, USA}
\author{T.~Holmstrom}
\affiliation{College of William and Mary, Williamsburg, Virginia 23187, USA}               
\author{C.E.~Hyde}
\affiliation{Old Dominion University, Norfolk, Virginia 23508, USA}
\affiliation{Universit\'e Blaise Pascal / CNRS-IN2P3, F-63177 Aubi\`ere, France}
\author{H.~Ibrahim}
\affiliation{Old Dominion University, Norfolk, Virginia 23508, USA}
\author{Y. Ilieva}
\affiliation{George Washington University, Washington D.C. 20052, USA}
\author{C.~W.~de~Jager}
\affiliation{Thomas Jefferson National Accelerator Facility, Newport News, Virginia 23606, USA}
\author{X.~Jiang}
\affiliation{Rutgers, The State University of New Jersey, Piscataway, New Jersey 08855, USA}
\author{M.~K.~Jones}
\affiliation{Thomas Jefferson National Accelerator Facility, Newport News, Virginia 23606, USA}
\author{H.~Kang}
\affiliation{Seoul National University, Seoul 151-747, Korea}
\author{A.~Kelleher}
\affiliation{College of William and Mary, Williamsburg, Virginia 23187, USA}               
\author{E.~Khrosinkova}
\affiliation{Kent State University, Kent, Ohio 44242, USA}
\author{E.~Kuchina}
\affiliation{Rutgers, The State University of New Jersey, Piscataway, New Jersey 08855, USA}
\author{G.~Kumbartzki} 
\affiliation{Rutgers, The State University of New Jersey, Piscataway, New Jersey 08855, USA}
\author{J.~J.~LeRose}
\affiliation{Thomas Jefferson National Accelerator Facility, Newport News, Virginia 23606, USA}
\author{R.~Lindgren}
\affiliation{University of Virginia, Charlottesville, Virginia 22904, USA}
\author{P.~Markowitz}
\affiliation{Florida International University, Miami, Florida 33199, USA}
\author{S.~May-Tal~Beck}
\affiliation{NRCN, P.O.Box 9001, Beer-Sheva 84190, Israel}
\author{E.~McCullough}
\affiliation{Saint Mary's University, Halifax, Nova Scotia B3H 3C3, Canada}
\author{D.~Meekins}
\affiliation{Thomas Jefferson National Accelerator Facility, Newport News, Virginia 23606, USA}
\author{M.~Meziane}
\affiliation{College of William and Mary, Williamsburg, Virginia 23187, USA}               
\author{Z.-E.~Meziani}
\affiliation{Temple University, Philadelphia, Pennsylvania 19122, USA}
\author{R.~Michaels}
\affiliation{Thomas Jefferson National Accelerator Facility, Newport News, Virginia 23606, USA}
\author{B.~Moffit}
\affiliation{College of William and Mary, Williamsburg, Virginia 23187, USA}               
\author{B.E.~Norum}
\affiliation{University of Virginia, Charlottesville, Virginia 22904, USA}
\author{Y.~Oh}
\affiliation{Seoul National University, Seoul 151-747, Korea}
\author{M.~Olson}
\affiliation{Saint Norbert College, Greenbay, Wisconsin 54115, USA}
\author{M.~Paolone}
\affiliation{University of South Carolina, Columbia, South Carolina 29208, USA}
\author{K.~Paschke}
\affiliation{University of Virginia, Charlottesville, Virginia 22904, USA}
\author{C.~F.~Perdrisat}
\affiliation{College of William and Mary, Williamsburg, Virginia 23187, USA}               
\author{E.~Piasetzky}
\affiliation{Tel Aviv University, Tel Aviv 69978, Israel} 
\author{M.~Potokar} 
\affiliation{Institute ``Jo\v{z}ef Stefan'', 1000 Ljubljana, Slovenia}
\author{R.~Pomatsalyuk}
\affiliation{Thomas Jefferson National Accelerator Facility, Newport News, Virginia 23606, USA}
\affiliation{Kharkov Institue, Kharkov 310108, Ukraine}
\author{I.~Pomerantz}
\affiliation{Tel Aviv  University, Tel Aviv 69978, Israel} 
\author{A.~Puckett}
\affiliation{Massachusetts Institute of Technology, Cambridge, Massachusetts 02139, USA}
\author{V.~Punjabi}
\affiliation{Norfolk State University, Norfolk, Virginia 23504, USA}
\author{X.~Qian}
\affiliation{Duke University, Durham, North Carolina 27708, USA}
\author{Y.~Qiang}
\affiliation{Massachusetts Institute of Technology, Cambridge, Massachusetts 02139, USA}
\author{R.~Ransome}
\affiliation{Rutgers, The State University of New Jersey, Piscataway, New Jersey 08855, USA}
\author{M.~Reyhan}
\affiliation{Rutgers, The State University of New Jersey, Piscataway, New Jersey 08855, USA}
\author{J.~Roche}
\affiliation{Ohio University, Athens, Ohio 45701, USA}
\author{Y.~Rousseau}
\affiliation{Rutgers, The State University of New Jersey, Piscataway, New Jersey 08855, USA}
\author{A.~Saha} 
\affiliation{Thomas Jefferson National Accelerator Facility, Newport News, Virginia 23606, USA}
\author{A.J.~Sarty}
\affiliation{Saint Mary's University, Halifax, Nova Scotia B3H 3C3, Canada}
\author{B.~Sawatzky}
\affiliation{University of Virginia, Charlottesville, Virginia 22904, USA}
\affiliation{Temple University, Philadelphia, Pennsylvania 19122, USA}
\author{E.~Schulte}
\affiliation{Rutgers, The State University of New Jersey, Piscataway, New Jersey 08855, USA}
\author{M.~Shabestari}
\affiliation{University of Virginia, Charlottesville, Virginia 22904, USA}
\author{A.~Shahinyan}
\affiliation{Yerevan Physics Institute, Yerevan 375036, Armenia}
\author{R.~Shneor}
\affiliation{Tel Aviv  University, Tel Aviv 69978, Israel} 
\author{S.~\v{S}irca}
\affiliation{Institute ``Jo\v{z}ef Stefan'', 1000 Ljubljana, Slovenia}
\affiliation{Dept. of Physics, University of Ljubljana, 1000 Ljubljana, Slovenia}
\author{K.~Slifer}
\affiliation{University of Virginia, Charlottesville, Virginia 22904, USA}
\author{P.~Solvignon}
\affiliation{Argonne National Laboratory, Argonne, Illinois 60439, USA}
\author{J.~Song}
\affiliation{Seoul National University, Seoul 151-747, Korea}
\author{R.~Sparks}
\affiliation{Thomas Jefferson National Accelerator Facility, Newport News, Virginia 23606, USA}
\author{R.~Subedi}
\affiliation{Kent State University, Kent, Ohio 44242, USA}
\author{S.~Strauch}
\affiliation{University of South Carolina, Columbia, South Carolina 29208, USA}
\author{G.~M.~Urciuoli}
\affiliation{INFN, Sezione di Roma, Piazzale Aldo Moro 2, 00185 Roma, Italy}
\author{K.~Wang}
\affiliation{University of Virginia, Charlottesville, Virginia 22904, USA}
\author{B.~Wojtsekhowski}
\affiliation{Thomas Jefferson National Accelerator Facility, Newport News, Virginia 23606, USA}
\author{X.~Yan}
\affiliation{Seoul National University, Seoul 151-747, Korea}
\author{H.~Yao}
\affiliation{Temple University, Philadelphia, Pennsylvania 19122, USA}
\author{X.~Zhan}
\affiliation{Massachusetts Institute of Technology, Cambridge, Massachusetts 02139, USA}
\author{X.~Zhu}
\affiliation{Duke University, Durham, North Carolina 27708, USA}
\collaboration{The Jefferson Lab Hall A Collaboration}
\noaffiliation
\date{\today}

\begin{abstract}
High-precision measurements of the proton elastic form factor ratio,
$\mu_{p} G_E^p/G_M^p$, have been made at four-momentum transfer, $Q^2$, values between 0.2 and 0.5 GeV$^2$.
The new data, while consistent with previous results,
clearly show a ratio less than unity and significant differences from the central values of several recent 
phenomenological fits.  By combining the new form-factor ratio data
with an existing cross-section measurement, one finds that in this $Q^2$ range
the deviation from unity is primarily due to $G_E^p$ being smaller than expected.
\end{abstract}


\maketitle

%\bibliographystyle{apsrev}

%%%%%%%%%%%%%%%%%% Main %%%%%%%%%%%%%%%%%%%%%%%%

Elastic scattering of electrons from protons reveals information 
about the distribution of charge and magnetism in the nucleon
via the electromagnetic form factors.
For decades, these form factors were determined by making Rosenbluth separations~\cite{rosenbluth} 
of cross-section results, as done for example in the reanalysis by Arrington~\cite{arr2004}.
Recently, however, high-quality polarized electron beams have  allowed polarization 
techniques~\cite{akhiezer1958,dombey1969,acg} to be used.
The new techniques revealed that the electric to magnetic proton form-factor ratio,
%, $\mu_p G_E^p/G_M^p$,
which was long thought to be nearly unity for all four-momentum transfers, $Q^2$,
becomes significantly less than unity at $Q^2 > 1$~GeV$^2$~\cite{jones00}.  This observation 
has led to a renewed experimental focus on the proton electromagnetic form 
factors~\cite{
gayou01,
pop2001,
%die2001,
gayou02,
%strauch03,
qattan05,
punjabi05,
jones06,
%bhu2006,
cra2007}.

A recent suggestion 
from a modern form-factor fit that there is structure in 
each of the four nucleon electromagnetic form factors for even 
$Q^2 < 1$~GeV$^2$ is intriguing~\cite{fandw2003} and has been discussed 
in recent review articles~\cite{hyde2004,perdrisat2006,arz2006}. 
The interest stems from the fact that changes of
just a few percent in the nucleon form factors at low $Q^2$ have
direct implications on our understanding of nucleon structure.
These include, but are not limited to, 
the weak form factors of the nucleon~\cite{acha2007,aniol1999,arm2005,maas2005}, 
generalized parton distributions accessed in DVCS~\cite{camacho2006},
generalized polarizabilities accessed in VCS near threshold~\cite{Roche:2000ng},
and the extraction of the Zemach radius~\cite{sick2003}. 

The highest precision data set of the ratio $\mu_p G_E^p/G_M^p$ at low $Q^2$, 
prior to the results reported herein, is 
from Bates BLAST~\cite{cra2007}.  This set has two out of 
eight points 2$\sigma$ (statistical) below unity with the average of the eight
points equal to 0.99 $\pm$ 0.01.  However, when systematic uncertainties are
included, no point is significantly lower than 1$\sigma$ from unity
and thus it was concluded~\cite{cra2007} that the data were consistent 
with unity.

%Such polarization data are available from 
%Jefferson Lab Hall A~\cite{jones00,punjabi05,gayou01,strauch03,bhu2006}, 
%Bates BLAST~\cite{cra2007},
%Mainz A1~\cite{pop2001,die2001}, and 
%Bates OHIPS~\cite{mil1998}. 

%Several recent fits~\cite{fandw2003,arr2004,kel2004,arr2006cfe,belushkin2006} are available,
%as well as a number of calculations.
%The fits (and the calculations) typically show a smooth fall-off of the form-factor ratio 
%with $Q^2$ that is in qualitative agreement with the data, but the fits
%differ from each other and from the data by up to a few percent.
%The differences arise from several sources, such as 
%which data are included in the fits, 
%treatment of higher-order effects in the cross section related to multiple-photon exchange, 
%such as Coulomb corrections, and
%constraints from the choice of particular functional forms.
%While the fits do not include the final BLAST data~\cite{cra2007},
%their variation indicates an improved data base is needed.  Note that 
%since the parameterizations do not include the final BLAST data, and furthermore,
%since some of the parameterizations rely on Rosenbluth separation of the form factors,
%differences between the parameterizations and high-precision polarization data
%should not be surprising.

In this work we present new, high precision measurements of $\mu_p G_E^p/G_M^p$ at $Q^2$ between 0.2 and 0.5 GeV$^2$
via the polarization transfer reaction $^1$H$(\vec{e},e'\vec{p})$.  
In the Born approximation the ratio
of the transferred transverse to longitudinal polarization relates to the electromagnetic
form factors by the equation:
\begin{equation}
R\equiv\mu_p\frac{G^P_E}{G^P_M}=-\mu_p \frac{E_e+E_e'}{2M}\tan\left(\frac{\theta_e}{2}\right)\frac{P_T}{P_L},
\label{eq:polxfer}
\end{equation}
where $\mu_p$ is the proton magnetic moment, $M$ is the mass of the proton, $E_e$ ($E_e'$) is the incident (scattered) electron energy,
$\theta_e$ is the electron scattering angle and $P_T$ ($P_L$) is the recoil proton polarization 
transverse (longitudinal) to the proton momentum.   In this approximation the third or normal polarization
component is zero. 

The experiment was 
performed in Hall A of the Thomas Jefferson National Accelerator Facility.
The polarized electrons, $\vec{e}$, were produced from a 
strained-superlattice GaAs crystal 
from the photoelectron gun~\cite{stutzman2007} 
and were accelerated to either 362 or 687~MeV.
The beam helicity state was flipped pseudo-randomly at 30 Hz;
beam charge asymmetries between the two helicity states were negligible. 
Due to multi-hall running the degree of longitudinal polarization in Hall A was limited
to 40\% rather than the full 80\%.
Note from Eq.~\ref{eq:polxfer} that $R$ is independent of the beam polarization, though the
uncertainties do increase as a result of the lower beam polarization.

The polarized beam was incident on a 15~cm long, liquid hydrogen target.
The kinematics of the measurements are given in Table~\ref{tab:kin}.
In all cases, the elastically scattered protons were detected in the left High Resolution Spectrometer, HRS,
which contains a Focal Plane Polarimeter, FPP. 
Six of the eight measurements were done as single-arm proton measurements, since
obstructions in the Hall prevented detecting electrons at angles larger than 60$^{\circ}$.
In the two measurements where it was possible, the coincident scattered electrons 
were detected in the right HRS. 
Details of the standard Hall A equipment can be found in~\cite{hallanim}.

\begin{table}[ht]
\begin{center}
\caption{\label{tab:kin} Kinematics and FPP
parameters for the measured data points.
The central spin precession angle is $\chi$. $\theta^{~p}_{lab}$ and $T_p$
are the proton lab angle and proton kinetic energy, respectively. S(C) 
denotes a single-arm (coincidence) measurement.  The analyzer material was 
carbon with a density $\approx$1.7~g/cm$^3$.
}
\begin{tabular}{|c|c|c|c|c|c|c|}
\hline
 $Q^2$ & $E_e$ & $\theta^{~p}_{lab}$ & $T_p$ & Analyzer Thickness& $\chi$& S/C \\
(GeV$^2$) & (GeV) & (deg) & (GeV) & (inches) & (deg) & \\

\hline
0.225 & 0.362 & 28.3 & 0.120 & 0.75  & 91.0& S\\
0.244 & 0.362 & 23.9 & 0.130 & 0.75  & 91.9& S\\
0.263 & 0.362 & 18.8 & 0.140 & 0.75  & 92.7& S\\
0.277 & 0.362 & 14.1 & 0.148 & 0.75  & 93.4& S\\
0.319 & 0.687 & 47.0 & 0.170 & 2.25  & 95.3& C\\
0.356 & 0.687 & 44.2 & 0.190 & 3.75  & 97.0& C\\
0.413 & 0.687 & 40.0 & 0.220 & 3.75  & 99.6& S\\
0.488 & 0.687 & 34.4 & 0.260 & 3.75  & 103.0& S\\
\hline
\end{tabular}
\end{center}
\end{table}

%The scattering angles, momentum, and interaction position
%at the target of each event were calculated from trajectories measured with
%vertical drift chambers~\cite{fissum2001}, located in the HRSs.
%Two planes of plastic scintillators provided triggering and time-of-flight 
%information for particle identification.
%The FPP, as typically configured~\cite{punjabi05}, consisted of two front and two rear straw chambers
%that determine the scattering of particles in a variable thickness carbon
%analyzer, with density $\approx$1.7 g/cm$^3$. 

For the singles data, it was necessary to apply cuts on the target interaction 
position, and to subtract residual end-cap events using spectra taken
on an aluminum dummy target.
The two coincidence points were essentially background free, due to the
large $ep$ cross section.
Quasi-elastic events from the target end-caps, through the Al($e,e^{\prime}p$)
reaction, were suppressed by requiring hydrogen elastic kinematics.

For the scattered protons, the polarization precesses as the particle
is transported through the spectrometer.   At the FPP, 
the transverse polarization components lead to azimuthal asymmetries in the 
re-scattering in the analyzing material due to spin-orbit interactions.
The alignment of the FPP chambers was determined with straight-through 
trajectories, with the analyzing material removed.
While misalignments and detector inefficiencies lead to false asymmetries, 
these false asymmetries largely cancel in forming the helicity differences
which determine the polarization-transfer observables.
The transferred polarization was determined by a 
maximum likelihood method using the difference of the 
azimuthal distributions corresponding to the two beam helicity states.
The spin transport in the spectrometer was taken into account
using a magnetic model calculation.
Previous Hall A measurements of the form-factor ratio
used the same procedures~\cite{jones00,gayou02,punjabi05,gayou01,strauch03,bhu2006}.

The form-factor ratio is determined from the ratio of 
polarization transfer components, and thus from the 
phase shift of the azimuthal scattering distribution in
the FPP analyzer. The analyzing power, efficiency, and beam polarization 
cancel out in the calculation of the form-factor ratio -- although they affect the size of the uncertainty;
thus, the main issue for systematic uncertainties is
spin transport in the spectrometer. 
The spin transport systematic uncertainties are determined by studying how the
form factor ratio changes when parameters such as reconstructed angles and
the spectrometer bend angle are changed by their uncertainties.
Detailed optical studies were performed to constrain the spin transport
for the first Hall A $G_E^P$ experiment~\cite{jones00}, which had the FPP mounted in HRS-right. 
The FPP was moved to HRS-left for the second $G_E^P$ experiment~\cite{gayou02} and has remained there
for subsequent experiments, but no similarly detailed optical studies have been performed. 
As the spectrometers are nearly identical, it is expected that the limiting systematic 
uncertainties in this measurement are similar, 
%These uncertainties were 
%about 0.4\% at $Q^2$ = 0.5 GeV$^2$, and rise with $Q^2$ in the range of $Q^2$ covered in~\cite{punjabi05}.
though since we lack the optical studies for HRS-left, our estimated
systematic uncertainties are twice as large
%, as can be seen from the
%point at ~0.5 GeV$^2$ in Table~\ref{tab:res}.

%The most detailed study of systematics was done for the first Hall~A $G_E^P$
%experiment~\cite{punjabi05}, which had the FPP mounted in HRS-right.
%As the spectrometers are nearly identical, it is expected that the
%limiting systematic uncertainties in this measurement are similar.
%These uncertainties were about 0.4\% at $Q^2$ $=$ 0.5 GeV$^2$, and 
%rise with $Q^2$ in the range of that experiment.
%Since we lack these detailed optical studies for HRS-left, our estimated
%systematic uncertainties are about twice as large.

To control the systematics in this experiment, each polarization point was measured at three
different spectrometer momentum settings, spaced 2 -- 3\% apart.
In all cases, the polarization values extracted were consistent
for the three settings. The uncertainties resulting from the subtraction 
of residual Al end-cap events were negligible compared with the 
other systematic uncertainties. The kinematics of the reaction are well determined by the 
recoil proton, thus there is no discernible improvement in the uncertainties when performing a 
coincidence measurement.
The benefit of the coincidence trigger is the suppression of background events, which
for a fixed data-acquisition rate allowed for higher statistics within a shorter time.

\begin{table}[ht]
\begin{center}
\caption{\label{tab:res} Shown are 
the experimental ratio results with statistical and systematic uncertainties 
along with
the FPP analyzing power $\langle A_C \rangle$ and efficiency $\varepsilon_{FPP}$ 
for a secondary scattering angle range of 5 to 40 degrees.
}
\begin{tabular}{|c|c|c|c|c|}
\hline
 $Q^2$ & $\langle A_c \rangle$ & $\varepsilon_{FPP}$ & FOM & $R \pm stat. \pm sys.$ \\
(GeV$^2$) &  & (\%) & (\%) & \\
\hline

0.225 & 0.16 & 1.17 & 0.03 &  0.9570 $\pm$ 0.0857 $\pm$ 0.0036\\
0.244 & 0.22 & 1.03 & 0.05 &  0.9549 $\pm$ 0.0500 $\pm$ 0.0037\\
0.263 & 0.24 & 1.04 & 0.06 &  1.0173 $\pm$ 0.0495 $\pm$ 0.0035\\
0.277 & 0.30 & 1.00 & 0.09 &  1.0060 $\pm$ 0.0504 $\pm$ 0.0030\\
0.319 & 0.34 & 6.05 & 0.70 &  0.9691 $\pm$ 0.0143 $\pm$ 0.0058\\
0.356 & 0.36 & 6.94 & 0.90 &  0.9441 $\pm$ 0.0099 $\pm$ 0.0050\\
0.413 & 0.46 & 4.73 & 1.00 &  0.9491 $\pm$ 0.0138 $\pm$ 0.0053\\
0.488 & 0.46 & 4.73 & 1.00 &  0.9861 $\pm$ 0.0189 $\pm$ 0.0094\\
\hline
\end{tabular}
\end{center}
\end{table}

The experimental results are summarized in Table~\ref{tab:res}.
The average FPP analyzing power $\langle A_c \rangle$ and efficiency $\varepsilon_{FPP}$ are consistent with
parameterizations of earlier FPP results~\cite{mcn}. 
The Hall A FPP design allows a much broader angular acceptance than many
previous devices, usually limited to about 20$^{\circ}$,
which leads to a slightly larger efficiency.
Also, at the lowest energies, the analyzing power increases at 
angles beyond 20$^{\circ}$, leading to a somewhat larger average
analyzing power.
The analyzing power quoted is the r.m.s. result, so that the FPP 
figure of merit, FOM, is given by $\varepsilon_{FPP}\langle A_c \rangle ^2$. 

\begin{figure}[ht]
\begin{center}
\includegraphics[width=3.5in]{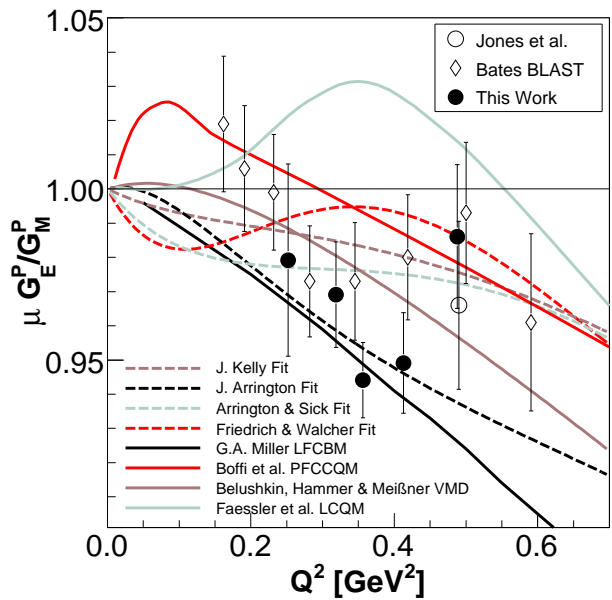}
\caption[]{\label{fig:data} (Color online) The proton form factor ratio
as a function of four-momentum transfer $Q^2$ shown with world data
with total uncertainties below 3\%~\cite{jones00,cra2007}.
The dotted and dash-dotted lines are fits~\cite{fandw2003,arr2004,kel2004,arr2006cfe}, while
the dashed and solid lines are from a vector-meson dominance calculation~\cite{belushkin2006},
light-front cloudy-bag model calculation~\cite{mil2002}, a light-front quark model calculation~\cite{faessler06},
and a point-form chiral constituent quark model calculation~\cite{boffi2002}.}
%The fits neglect two-photon effect corrections except for that by Arrington \& Sick~\cite{arr2006cfe} 
%which includes Coulomb corrections.}
\end{center}
\end{figure}

The new data, along with other high precision results, are shown in Fig.~\ref{fig:data} with
the four data points taken at 362~MeV beam energy
having been combined into a single point for plotting.  Included in the figure are 
a representative sample of the numerous modern calculations and fits that are avaliable.  
%The fits neglect two-photon effect corrections, although that by
%Arrington \& Sick~\cite{arr2006cfe} include Coulomb distortions.
The high statistical precision points at $Q^2$ = 0.356 and 0.413 GeV$^2$ clearly indicate 
that $R < 1$. 
While the BLAST data alone were consistent with unity~\cite{cra2007}, usually at the upper
end of the uncertainty,
the BLAST data are also consistent with the new measurements, and the combination of the two 
data sets is clearly not consistent with unity.
The point at 0.356~GeV$^2$ is 5$\sigma$ (stat. + syst.) below unity and
the point at 0.413~GeV$^2$ is 3.4$\sigma$ below unity; 
previous data were within $\sim$2$\sigma$(stat.) of unity.

Although a smooth fall-off of $\mu_p G_E^p/G_M^p$ with $Q^2$ is not ruled out,
the new data hint at a local minimum in the form-factor ratio at about 0.35 -- 0.4
GeV$^2$.  Assuming uncorrelated uncertainties, in the range $Q^2$ = 0.3 -- 0.45 GeV$^2$,
we find the world data including the current work average to $0.960\pm0.005\pm0.005$.
This is $3\sigma$ lower than the neighboring $Q^2$ range 0.45 -- 0.55 GeV$^2$, where $R = 
0.987\pm0.005\pm0.006$.
In this latter range, the form-factor ratio is only $1.6\sigma$ below unity.
Calculations which tend to agree with the new ratio results, such as the light-front cloudy bag 
model calculation by G.A.~Miller~\cite{mil2002},
%and the light-front constituent quark model by Cardarelli {\it et al.}~\cite{card2000}, 
however, show a monotonic decrease of the form-factor ratio.  Additional calculations may be found in~\cite{punjabi05}.

\begin{figure}[ht]
\begin{center}
\includegraphics[width=3.5in]{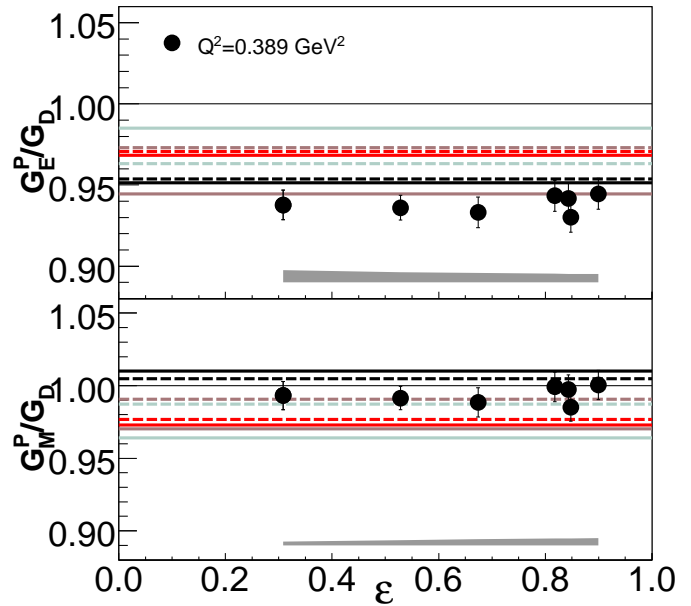}
\caption[]{\label{fig:FFs} (Color online) The extracted individual
proton form factors as a function of $\varepsilon$.  The form factors were obtained for a single $Q^2$ 
value using the average of the 0.356 and 0.413~GeV$^2$ data of this work and existing cross-section data 
at 0.389~GeV$^2$~\cite{berger1971}.
The error bars indicate the statistical error of the Berger et al. data while the shaded 
region indicates how the uncertainty on the asymmetry shifts the points.
The systematic uncertainty of the cross section experiment, approximately 2\% on each form factor, 
have not been included.
The lines are the same as in Fig.~\ref{fig:data}.}
\end{center}
\end{figure}

By combining the present measurement with previous cross-section results, it is
possible to extract the individual form factors.  
This was done by combining the highest precision existing
cross-section data in the vicinity of the measured ratio~\cite{berger1971} at $Q^2=0.389$ GeV$^2$ with the average of our 
form-factor ratios from $Q^2=0.36$ GeV$^2$ and $0.41$ GeV$^2$.
Figure~\ref{fig:FFs}, which uses the same codes as Fig.~\ref{fig:data}, shows that the form-factor 
extraction is essentially independent of $\varepsilon$, the virtual photon polarization, over the extracted range.  
Interestingly, the deviation 
from unity in the ratio seems to be dominated by the electric form factor. 
This result 
is consistent with previous Rosenbluth separation measurements and fits
in this region of $Q^2$; the Rosenbluth results tend to have $\sim$1--3\%
uncertainties for each of the form factors, while the fits vary by
several percent for each~\cite{arr2004}. 
While the Belushkin {\it et al.} calculation~\cite{belushkin2006} generally fits best over the full $Q^2$
range of this measurement, and at $Q^2=0.389$ GeV$^2$ is closest to $G_E^p$, it 
over-predicts $R$ by underestimating $G_M^p$.
The best fit of the ratio at $Q^2=0.389$ GeV$^2$ is from Arrington~\cite{arr2004}, which
over-predicts each form factor by 1-2\%.
The Miller calculation predicts the ratio at Q$^2$ = 0.389~GeV$^2$ well,  
but also over-predicts each form factor by about 1-2\%.  
In fact, none of these modern calculations predicts both the individual form factors 
and the ratio correctly.  Some calculations, which were not shown, such as the light-front constituent quark model
by the Cardarelli {\it et al.}~\cite{card2000} are in good agreement with the
form-factor ratio data in this $Q^2$ range, but the individual form factors are significantly overestimated
by present quark potential models~\cite{simulapriv}.

The comparison of fits with the new data
%our new results along with the new BLAST results, 
suggests that a critical reexamination is needed of 
experiments (e.g.~\cite{ncg2006,arm2005,camacho2006}) that require a knowledge of
low $Q^2$ form factors to a precision of better than $\sim$3\%.
For example, for the HAPPEx measurement of the weak form factors~\cite{aniol2004long} the new data 
adjust the measured asymmetry by about -0.5~ppm, corresponding to a smaller effect from strange quarks,
on data with a statistical uncertainty of $\approx$1~ppm.
More significantly,
this new result would shift the expected HAPPEx-III result~\cite{paschke2005} by one standard deviation.  

%Finally, the low Q$^2$ proton form-factor database is likely to be improved in the 
%next few years, which should give a clear answer to whether there are few
%percent structures in the separated form factors or in the ratio $\mu_p G_E^p/G_M^p$.
%Upcoming results include new cross sections from Mainz~\cite{mainz} and
%from Hall~A~\cite{ledexed}.  The Hall~A cross-section results will be
%for the same $Q^2$ as reported herein, and thus will allow a more direct extraction of
%individual form factors.
%In addition, the current work is the result of FPP measurements that were run 
%for limited times with limited beam polarization.
%A dedicated experiment~\cite{ron2007}
%could easily improve the statistical uncertainties by about 
%a factor of three, with normal Jefferson Lab beam polarization and day-long runs,
%leading to a much improved database for $\mu_p G_E^p/G_M^p$. The results of this work indicate 
%the importance of cross-checking cross-section measurements with high-precision ratio 
%measurements from polarization techniques.

In summary, we made polarization-transfer measurements to precisely determine
the proton form-factor ratio at low $Q^2$.
We showed that the form-factor ratio 
differs from unity at low $Q^2$ and that the deviation is most likely
dominated by the electric form factor. 
Our data suggest a lower value of the ratio and electric form factor than
many modern fits. 
No fit or calculation adequately represents the ratio and extracted form factor 
data over the entire range.
%A more definitive measurement is both relatively easy and highly desirable, given the 
%possible implications~\cite{ron2007}.

We thank the Jefferson Lab physics and accelerator divisions for their support.
This work was supported by 
the U.S.\ Department of Energy,
the U.S.\ National Science Foundation,
Argonne National Laboratory under contract DE-AC02-06CH11357,
the Israel Science Foundation, the Korea Research Foundation, the US-Israeli Bi-National Scientific
Foundation, and the Adams Fellowship Program of the
Israel Academy of Sciences and Humanities.  
Jefferson Science Associates operates
the Thomas Jefferson National Accelerator Facility under DOE
contract DE-AC05-06OR23177.
The polarimeter was funded by the U.S.\ National Science
Foundation, grants PHY 9213864 and PHY 9213869.

%\bibliography{ledex}

\end{document}